\documentclass[aps,prb,groupedaddress,showpacs,preprint,showkeys]{revtex4}
\usepackage{graphicx,natbib,amssymb,amsmath,color,subfigure}
\begin{document}

\title{Effect of Ir substitution in the ferromagnetic superconductor
RuSr$_{2}$GdCu$_{2}$O$_{8}$}

\author{A. R. Jurelo$^{1}$}
\altaffiliation[Permanent address: ]{Universidade Estadual de Ponta Grossa, Ponta Grossa,
Paran$\acute{a}$, Brazil.}
\author{S. Andrade$^{1}$, R. F. Jardim$^{1}$}\email{rjardim@if.usp.br}
\author{F. C. Fonseca$^{2}$}
\author{M. S. Torikachvili$^{3}$}
\author{A. H. Lacerda$^{4}$}
\author{L. Ben-Dor$^{5}$}

\affiliation{$^{1}$Instituto de F\'{i}sica, Universidade de S\~ao
Paulo, CP 66318, 05315-970, S\~ao Paulo, SP, Brazil \\
$^{2}$Instituto de Pesquisas Energ\'eticas e Nucleares, CP 11049,
05422-970, S\~ao Paulo, SP, Brazil \\ $^{3}$Department of
Physics, San Diego State University, San Diego, California 92182-1233 \\
$^{4}$National High Magnetic Field Laboratory (NHMFL), Los Alamos
National Laboratory, Los Alamos, New Mexico 87545 \\
$^{5}$Department of Inorganic and Analytical Chemistry, Hebrew University, Jerusalem
91904, Israel}

\date{\today}

\begin{abstract}

A detailed study of the effect caused by the partial substitution
of Ru by Ir on the magnetic and superconducting properties of the
ruthenocuprate Ru$_{1-x}$Ir$_{x}$Sr$_{2}$GdCu$_{2}$O$_{8}$; 0
$\leq$ \emph{x} $\leq$ 0.10; is presented. The combined
experimental results of structural, electrical, and magnetic
measurements indicate that Ir substitutes Ru for $x\leq$ 0.10 with
no significant structural distortions. Ir-doping gradually
suppresses both the magnetic and the superconducting states.
However, all samples were observed to attain the zero-resistance
state at temperatures $\geq$ 2 K up to the highest applied
magnetic field of 18 T. The resistive upper-critical field
\emph{H$_{c2}$} as a function of temperature has been determined
for these polycrystalline samples. Values of \emph{H$_{c2}$}(0)
were found to be $\thicksim$ 52 T, and weakly dependent on the Ir
concentration. We have also observed that the superconducting
transition width decreases and the slope of the resistive
transition increases with increasing Ir doping, a feature which is
much more pronounced at high applied magnetic fields. The
double-peak structure observed in the derivative of the resistive
curves has been related to an inhomogeneous nature of the physical
grains which is enhanced due to the Ru substitution by Ir. This
indicates that the Josephson-junction-array (JJA) model seems to
be appropriated to describe the superconducting state in these
ruthenocuprates. The low temperature \emph{$\rho(T)$} data along
with the determined vortex thermal activation energy are
consistent with a 2D vortex dynamics in these materials.

\end{abstract}



\maketitle

\section{Introduction}

The study of the coexistence of superconductivity and magnetic
ordering in the ruthenocuprate RuSr$_{2}$GdCu$_{2}$O$_{8}$
(Ru-1212) has attracted great interest since the original study of
Bauernfeind \emph{et al.}.\cite{1} The Ru-1212 is a 1212-type
layered cuprate structurally similar to the
YBa$_{2}$Cu$_{3}$O$_{7-\delta}$(YBCO), where Y and Ba are replaced
by Gd (or Eu) and Sr, respectively, and the Cu-O chains replaced
by RuO$_{2}$ layers.\cite{1,2} In these materials the magnetic
long-range ordering of the Ru sub-lattice occurs below a
transition temperature \emph{T$_{M}$} $\sim$ 130 K while the
superconductivity arising from the CuO$_{2}$ layers occurs below a
critical temperature \emph{T$_{c}$} $\sim$ 40 K. Recent studies
have indicated that the magnetic ordering and the superconducting
state are essentially decoupled, being related only by the charge
transfer between Ru and CuO planes.\cite{3} In addition to this,
there is no clear evidence about the exact nature of the magnetic
structure of these materials up to now but it is accepted that,
for low magnetic fields, there is an antiferromagnetic (AFI)
order, whereas for high magnetic fields ($\emph{H}$ $\backsim$ 2
T) a spin-flop transition is observed,\cite{4} with Ru magnetic
order essentially ferromagnetic (FM). Also, from experimental data
and theoretical analysis, it was proposed that for temperatures
lower than \emph{T$_{c}$}, the magnetic-flux lines are present
even without an external magnetic field, suggesting the creation
of a spontaneous vortex phase (SVP).\cite{5,6}

In fact, the genuine coexistence of superconductivity and
magnetism at microscopic level is still controversial. Some
experimental studies\cite{7} have shown that the ruthenocuprates
are microscopically uniform. On the other hand, several
experimental results have indicated a possible phase-separation of
superconducting (SC) and magnetic regions.\cite{8,9,10} For
instance, high-resolution transmission electron microscopy (HRTEM)
and synchrotron X-ray diffraction analysis have suggested a phase
separation in Ru-1212 compounds.\cite{8} It was argued that such a
phase separation arises from the rotation of the RuO$_{6}$
octahedra around the $\emph{c}$-axis, resulting in the formation
of small domains with characteristic lenghts $\leq$ 200 {\rm\AA}
separated by sharp antiphase boundaries of reversed
rotations.\cite{8} Also, a phase separation between FM and AFM
nanodomains inside physical grains of Ru-1212 has been proposed
from a detailed analysis of magnetization data.\cite{9,10} The
authors have concluded that intragrain properties of the
ruthenocuprates exhibit features of granular superconductors and a
Josephson-junction-array (JJA) model was invoked to account for
the intrinsic inhomogeneities of intragrain
superconductivity.\cite{9,10} Therefore, a discussion of whether
both superconducting and magnetic phases originate from the same
crystallographic structure, and features of this intimate
coexistence on a microscopic scale are relevant questions for the
understanding of these materials.

Considering that the coupling allowing for the coexistence of
superconductivity and ferromagnetism in Ru-1212 compounds is very
weak and strongly affected by chemical substitutions, the dilution
of the magnetic Ru sublattice by different ions is an interesting
approach to probe the coexistent phenomena. The partial
substitution of Ru by Sn$^{4+}$ was found to suppress the FM
moment of the sublattice and to increase the onset of the SC
transition. These features would reflect an increase in the
transfer rate of holes to the CuO$_{2}$ planes.\cite{11} Studies
regarding substitution of Ru by both Ti and Rh revealed that both
FM and SC transition temperatures are reduced upon increasing
dopant concentration.\cite{12} The substitution of Ru by Nb$^{5+}$
results in a decrease of the magnetic ordering temperature and an
increase in the Ru valence,\cite{13} whereas for Ta-substituted
specimens an apparent suppression of the superconductivity of
Ru-1212 has been observed.\cite{14} In general, both magnetic and
superconducting properties of the ruthenocuprates are affected by
the ionic radius, valence, and magnetic character of the
substituting ion. However, changes observed in alloying Ru-1212
compounds are usually accompanied by significant structural
distortions due to differences in ionic radii. Within this
scenario it is a difficult task to distinguish between changes
arising from properties of the substituted ion and those from
crystallographic distortions.

In the present work we have investigated the crystallographic,
transport, and magneto-transport properties of
Ru$_{1-x}$Ir$_{x}$Sr$_{2}$Gd$_{1}$Cu$_{2}$O$_{8}$ compound in
order to study the relationship between superconductivity and
magnetism. We have found that Ir substitutes Ru up to 10 \% in
Ru-1212 without appreciable structural changes. In addition to
this, the combined data indicate a possible phase-separation in
the Ru-1212 compound.

\section{Experimental Procedure}\label{experimental}

Polycrystalline samples of
Ru$_{1-x}$Ir$_{x}$Sr$_{2}$GdCu$_{2}$O$_{8}$ (Ru(Ir)-1212);0 $\leq$
\emph{x} $\leq$ 0.10; were prepared following a two-step
procedure.\cite{15} The two-step synthesis minimizes the formation
of the SrRuO$_{3}$ phase, yielding samples with better
quality.\cite{16} Initially, the
Sr$_{2}$GdRu$_{1-x}$Ir$_{x}$O$_{6}$ (Sr-2116) precursor was
prepared by mixing stoichiometric quantities of high purity Ru,
Ir, SrCO$_{3}$, and Gd$_{2}$O$_{3}$, grinding together and heating
in air at 1250 $^\circ$C for 12 h. Then, CuO was mixed to the
Sr-2116 powders, ground together, pressed into pellets, and
sintered at 1060 $^\circ$C for 72 hours in flowing O$_{2}$. The
crystal structure of the samples was analyzed by X-ray powder
diffraction (XRD) measurements using CuK$_{\alpha}$ radiation on a
Bruker D8 Advance diffractometer. The diffraction patterns were
collected in the 2\emph{$\Theta$} range 20$^\circ$ to 80$^\circ$
with a step of 0.01 and 8 s counting time. Rietveld refinements of
crystal structures were performed using the GSAS software. The
temperature dependence of the magnetoresistance \emph{$\rho(H,T)$}
was measured by the standard four-probe method using a Linear
Research Model LR-700 bridge operating at 16 Hz. In all transport
measurements, copper electrical leads attached to Ag film contact
pads (made with Ag epoxy) on parallelepiped-shaped samples with
typical dimensions of 5 x 2 x 1.5 mm$^3$. The magnetoresistance
experiments were performed at the National High Magnetic Field
Laboratory, Los Alamos, in the temperature range from 2 to 300 K
and under magnetic fields $\emph{H}$ up to 18 T. Measurements at
low applied magnetic fields $\emph{H}$ up to 0.5 T were performed
in a home-made apparatus using a superconducting coil with very
low remnant field. The samples were characterized by both
magnetization $\emph{M(T)}$ and $\it{ac}$ magnetic susceptibility
$\emph{$\chi_{ac}$(T)}$ using a SQUID magnetometer from Quantum
Design. Magnetization measurements in the remnant field ($\sim$ 1
Oe) of the superconducting magnet were performed in the
temperature range from 5 to 300 K in both zero-field-cooled (ZFC)
and field-cooled (FC) modes. The T-dependence of the $ac$ magnetic
susceptibility (f = 155 Hz) was measured with an excitation field
of 2 Oe.

\section{Results and Discussion}\label{exresult}

The X-ray diffraction patterns of
Ru$_{1-x}$Ir$_{x}$Sr$_{2}$GdCu$_{2}$O$_{8}$ (Ru(Ir)-1212) for
\emph{x} = 0.00 and \emph{x} = 0.10 are displayed in Fig. 1. All
the samples were found to be nearly single-phase although small
fractions of SrRuO$_{3}$ and Sr-2116 ($\leq$ 2 \%) could be
detected. The volume fraction of the extra phases were found to
show no dependence on the Ir concentration. The diffraction peaks
of the desired phase were indexed as belonging to the Ru-1212
tetragonal phase, space group \textit{P4/mmm}. The inset of Fig. 1
shows the calculated lattice parameters as a function of the Ir
content in this series. The refined lattice parameters of the
pristine compound \emph{a} = \emph{b} = 3.8389(1) $\rm\AA$, and
\emph{c} = 11.5652(1) $\rm\AA$ are close to the values reported
previously.\cite{8} The six-fold coordination of Ir$^{5+}$ and
Ir$^{4+}$ have ionic radii values very close to Ru$^{5+}$ and
Ru$^{4+}$,\cite{17} respectively, and the lattice parameters
yielded by the Rietveld analysis are essentially Ir content
independent, as displayed in Fig. 1. In addition, the Rietveld
refinement for the specimen with Ir \emph{x} = 0.10 doesn't show
an orthorhombic distortion, which further suggests that the space
group \textit{P4/mmm} is preserved for the range of Ir content
here investigated.\cite{8} This assumption is supported by recent
results where the Ir-1212 phase has been successfully synthesized
and found to exhibit a tetragonal crystal structure, space group
\textit{P4/mmm}.\cite{18} These results indicate that Ir
substitutes Ru in this series.  The crystallographic parameters
for the samples with \emph{x} = 0.00 and 0.10 specimens obtained
by the refinements are summarized in Table I.

Figure 2 shows the temperature dependence of the electrical
resistivity of Ru(Ir)-1212 at zero external magnetic field. The
measured $\rho$(T) are within the range of the reported values for
the Ru-1212 compound and, upon Ir doping the resistivity at room
temperature increases from $\rho$(300 K) $\sim$ 15 m$\Omega$cm to
$\sim$ 21 m$\Omega$cm for \emph{x} = 0.00 and 0.10,
respectively.\cite{8,19} Both the pure and Ir-substituted samples
exhibit metallic behavior in the normal state, and the value of
$d\rho/dT$ is higher in Ir-doped compounds. Subtle falls in the
$\rho$(T) curves near 130 K correlate well in temperature with the
onset of magnetic ordering, and are probably related to the
suppression of the spin-flip scattering. This feature is clearly
observed as a maximum in the \emph{d$\rho$/dT} data near the
magnetic ordering temperature \emph{T$_{M}$} $\sim$ 130 K for
\emph{x} = 0.00, as shown in the inset of Fig. 2. Below
\emph{T$_{M}$}, the $\rho$(T) curves of Ru(Ir)-1212 specimens
exhibits an extended metallic region, which is consistent with a
simple two-band model proposed recently.\cite{20} The electrical
resistivity displays a shallow minimum near 65 K for the sample
with \emph{x} = 0.00, and a more pronounced one at 72 K for the
compound with \emph{x} = 0.10. The $\rho$(T) minimum is followed
by a slight upturn in $\rho$(T) close to the onset of the
superconductivity (\emph{T$_{c,onset}$}). This upturn is hardly
seen in samples with low Ir content; however it becomes
discernible in the specimen with \emph{x} = 0.10. In addition, the
\emph{T$_{c,onset}$} decreases from $\sim$ 50 K for \emph{x} =
0.00 to $\sim$ 30 K for \emph{x} = 0.10.

These features may be related to the electronic mean free path
which can be extracted from the \emph{$\rho$} according
to\cite{21}

\begin{equation}
\\l = \frac{(4.95\texttt{x}10^{-4})v_{F}}{(\hbar w_{p})^{2} \rho}, \label{eq1}
\end{equation}

\noindent where \emph{v$_{F}$} is the Fermi velocity, assumed to
be 2.5 x 10$^{-7}$ cm/s, and \emph{$\rho$} is given in
$\mu$$\Omega$cm.\cite{22,23} Considering \emph{$\hbar$w$_{p}$}
$\sim$ 0.2 eV,\cite{21} the values of $\emph{l}$ at 300 K were
calculated using Eq.(1) and are summarized in Table II. For the
sample with \emph{x} = 0.00, $\emph{l}$ = 21 $\rm\AA$ a value that
decreases with increasing Ir concentration, reaching $\emph{l}$ =
15 $\rm\AA$ for the \emph{x} = 0.10 composition. For the Ru-1212
compound, $\emph{l}$ is over 10 times larger than the typical Cu-O
bond length ($\sim$ 1.9 \rm\AA) in these materials. In addition,
this value is lower than that observed for Ru-1222 (close to 58
\rm\AA)\cite{24} but comparable to the ones found in other
high-\emph{T$_{c}$} cuprates.\cite{21} One can also infer from the
$\rho$(T) data that increasing Ir content results in a systematic
decrease of the superconducting transition temperature and an
increase of the normal-state electrical resistivity. These
features may have their counterpart in the magnetic properties of
these Ru(Ir)-1212 compounds, as discussed below.

The ZFC (open symbols) and FC (full symbols) measurements of
magnetization in Fig.3 were taken in the remnant field of the
superconducting magnet (\emph{H}$\sim$ 1 Oe) for samples with Ir
content \emph{x} = 0, 0.05, and 0.10. At temperatures below 30 K,
these compounds exhibit superconductivity and diamagnetic
contributions are clearly observed in both ZFC and FC curves. On
the other hand, no diamagnetic signature has been observed for
$\emph{M(T)}$ measurements in very low applied magnetic fields, as
low as $H$ = 5 Oe (data not shown). The absence of appreciable
diamagnetism in low applied fields is a common feature of these
Ru-1212 compounds. The strength of the diamagnetic response is
strongly dependent on sample preparation and, consequently, on the
relative volume fraction of superconducting and
non-superconducting phases due to the so-called spontaneous vortex
phase, which arises even in zero applied magnetic field.\cite{5}
For temperatures below 15 K, a positive upturn in the FC
magnetization curve associated with the paramagnetic contribution
of the Gd$^{3+}$ ions has been observed. Such a feature is in
excellent agreement with heat capacity measurements (data not
shown) that suggest the development of antiferromagnetic ordering
below $T$ $\sim$ 2.5 K at the Gd sublattice for samples with
\emph{x} = 0.00 and \emph{x} = 0.10. However, we have observed
that the Neel temperature of the Gd sublattice is not modified by
the Ir substitution.  This further indicates that Ir
preferentially substitutes Ru in this series.

In the upper inset of Fig. 3, the magnetic ordering transition of
the Ru-sublattice is inferred from a pronounced peak in the
in-phase component of the $\emph{$\chi_{ac}$(T)}$ magnetic
susceptibility data. A careful inspection of the figure indicates
that $\emph{$\chi_{ac}$(T$_{M}$)}$ = 130 K for the sample with
\emph{x} = 0.00. The magnetic transition temperature
\emph{T$_{M}$} is very sensitive to the Ir concentration and
decreases to \emph{T$_{M}$} $\sim$ 112 K for the sample with
\emph{x} = 0.10. Also, from the upper inset of Fig. 3, one is able
to infer that the magnitude of the magnetic moment decreases with
increasing Ir concentration. The latter result, combined with the
gradual decrease of \emph{T$_{M}$} with increasing Ir content, is
compelling evidence that Ir does replace Ru in this series. In the
lower inset of Fig. 3, \emph{T$_{M}$} as a function of Ir content
is displayed. The results indicate that \emph{T$_{M}$} decreases
linearly with Ir content at the rate of $\sim$ -1.6 K/Ir at \%.
Values of \emph{T$_{M}$} are summarized in Table II.

The suppression of the diamagnetic signal in Ru(Ir)-1212 at very
low applied magnetic fields $\sim$ 5 Oe is of interest and has
been discussed previously for similar compounds.\cite{9} It was
argued that these oxides are comprised of two different phases due
to a phase separation phenomenon: one superconducting and another
one which is magnetic. These phases, which are believed to exist
in a nanoscale dimensions, are homogeneously distributed
throughout the material and coexist at low temperatures, a
morphology similar to granular superconductors.\cite{25} Within
this context, the absence of appreciable diamagnetism is a
consequence of the reduced dimensions of the superconducting
regions which are comparable to the large London penetration
depth, as discussed elsewhere.\cite{25} Thus, in order to clarify
the effects caused by the application of applied magnetic fields
on the superconducting properties of Ir-substituted Ru-1212
materials we have carried out magnetoresistivity
\emph{$\rho(H,T)$} measurements in applied magnetic fields up to
18 T. Few selected curves of \emph{$\rho(H,T)$} are shown in Fig.
4 for two samples with \emph{x} = 0.00 and 0.10. The
\emph{$\rho(H,T)$} curves for the sample with \emph{x} = 0.00
indicate that \emph{T$_{c,onset}$} remains nearly constant $\sim$
50 K under magnetic fields. On the other hand, the temperature in
which zero-resistance is attained (\emph{T$_{c,zero}$}) decreases
rapidly for low applied magnetic fields (\emph{H} $\leq$ 2 T),
followed by a much less pronounced drop in higher \emph{H}. These
features certainly resemble the ones observed in granular
superconductors.\cite{25} We mention that the $\rho$($T$, $H$ = 0)
curves displayed in Fig. 4 were taken in the remnant field of the
Nb$_{3}$Sn superconducting magnet, estimated to be $\sim$ 0.03 T.
Therefore, the transition width of $\rho$($T$) is broadened from
the expected one for the true zero-field data. The
\emph{$\rho(H,T)$} data also show that the sample with the lowest
\emph{T$_{c,onset}$} ($x$ = 0.10) attains the zero-resistance
state at $T$ $\geq$ 2 K even for the highest applied magnetic
field of 18 T. This result reinforces the picture of a granular
behavior in this series since pathways are still preserved within
the material even at 18 T. A detailed analysis of the low-field
data probing the granular properties of these materials is
described below.

In order to further probe the effect of the applied magnetic field
on the resistive transition, the derivative curves of the
electrical resistance versus temperature were constructed. For
example, displayed in Fig. 5 are the d$\rho$($H$,$T$)/d$T$ curves
for the Ir-doped sample with \emph{x} = 0.02 taken at several
applied magnetic fields. At \emph{H} = 0, a sharp peak at
\emph{T$_{1}$} $\sim$ 38 K is observed. The value of
\emph{T$_{1}$} depends on the concentration of Ir; for \emph{H} =
0, the transition temperature decreases from $\emph{T$_{1}$}$
$\sim$ 39 K ($\emph{x =}$ 0.00) to $\emph{T$_{1}$}$ $\sim$ 22 K
($\emph{x =}$ 0.10). The value of \emph{T$_{1}$} is nearly
magnetic field independent \emph{H} up to 0.5 T. However, two
features in the \emph{H} $\sim$ 0.035 - 0.500 range are of
interest: (i) the reduction in the peak intensity; and (ii) a
pronounced broadening of the peaks, culminating in a split into
two convoluted peaks. A further increase in \emph{H} $\geq$ 0.5 T
results in a progressive reduction of the amplitude of the high
temperature peak, leading to its gradual suppression, and the
appearance of another peak at low temperatures for applied
magnetic fields $\emph{H}$ $\sim$ 1 - 2 T. The amplitude of the
peak at lower temperatures increases strongly with \emph{H}, and
its position is monotonically shifted towards lower temperatures.
Similar double-peak structure and its evolution with \emph{H} were
observed for all samples studied. However, the
d$\rho$($H$,$T$)/d$T$ curves at high magnetic fields ($\emph{H}$
$\geq$ 2 T) reveal that the amplitude, width, and position of the
lower peak depend on the Ir content (see inset of Fig. 5). For
$\emph{H =}$ 14 T, the intensity of the peak increases and its
width decreases with increasing Ir concentration.

We also mention that similar magnetic field dependence of the
double peak behavior seen in d$\rho$($H$,$T$)/d$T$ curves has been
previously observed in both Ru-1212 and Ru-1222
compounds.\cite{24,25,26} The two-peak feature in
d$\rho$($H$,$T$)/d$T$ versus \emph{T} data is usually related to
the development of superconductivity within the grains
(intragrain) and between grains (intergrain) at an upper and lower
temperatures, respectively.\cite{25} However, it has been argued
that ruthenocuprates exhibit granular behavior, a feature
consistent with a phase separation of mesoscopic superconducting
and non-superconducting phases even within the grains.\cite{9,10}
In the present case one may consider that Ru ions are replaced by
Ir and this would affect preferably the intragrain properties, as
inferred from both $\emph{M(T)}$ and $\rho$(T) data. On the other
hand, the inset on Fig. 5 shows that Ir substitution changes the
shape of the low temperature peak in d$\rho$(H,T)/dT at high
applied \emph{H}. Such an observation suggests that the low
temperature peak at high \emph{H} may not be related only to the
so-called intergranular transition. In fact, as the studied
samples attain zero resistance state at \emph{T} $\geq$ 2 K up to
highest applied magnetic fields of 18 T, it is expected that the
intragrain transition takes place at magnetic fields within the
studied range. Our results evidence a strong-field dependence on
both the magnetic and transport properties of the Ru(Ir)-1212,
indicating that inhomogeneities are present within physical
grains.

Such a double-peak structure in d$\rho$($H$,$T$)/d$T$ versus
\emph{T} in ruthenocuprates has been previously observed and
discussed within the scenario of with weakly disordered
Josephson-junction arrays (JJA).\cite{9,10} It was argued that
nanoscale superconducting domains are coupled through Josephson
junctions below the thermodynamic transition temperature, as a
consequence of a phase separation into FM and AFI
regions.\cite{9,10} Our results seem to be consistent with such a
scenario also due to the fact that the high temperature peak
vanishes for applied magnetic fields $\emph{H}$ $\sim$ 1 - 2 T,
which is actually the same value where a spin flop-like transition
has been observed.\cite{4} Thus, it seems reasonable to consider
that even though granular behavior may be present, the
magnetoresistivity results of the Ir-substituted samples suggest
that the lower and high temperature peaks are a consequence of the
intragrain granular structure due to a phase separation.

The main effect of the applied magnetic field is the broadening of
the resistive transition due to the movement of vortices (see
Figs. 4 and 5). This result indicates the presence of dissipation
phenomena as commonly observed in conventional high-temperature
superconductors. Thus, by using the Arrhenius-type
expression\cite{27}

\begin{equation}
\\ \rho = \rho_{0}e ^{{-}\frac{U}{k_{B}T}}, \label{eq2}
\end{equation}

\noindent where \emph{$\rho_{0}$} is the order of
\emph{$\rho$}(300 K) and \emph{k$_{B}$} is the Boltzmann constant,
one is able to fit the low part of \emph{$\rho(H,T)$} curves to
obtain the vortex thermal activation energy \emph{U}. Fig. 6 shows
the Arrhenius plots of the resistive transitions for the sample
with \emph{x} = 0.05 in applied magnetic fields up to 18 T.
Pinning energies \emph{U} can be estimated from the slopes, over
which the data can be represented by a straight line, as shown by
solid lines in Fig 6. The same procedure has been adopted for
other samples and the \emph{U} values obtained for the \emph{x} =
0.05 sample, when the applied field increases from 0 to 18 T, are
\emph{U} = 60 meV and 5 meV, respectively. The inset of Fig. 6
shows the \emph{U} values for \emph{x} = 0.00 and \emph{x} = 0.05
as a function of the applied magnetic field. At \emph{H} = 0.005
T, as the Ir content increase from \emph{x} = 0.00 to \emph{x} =
0.05, the \emph{U} value decreases from 100 meV to 60 meV probably
due to the narrowing of the superconducting transition with
increasing Ir content. For higher applied fields, \emph{U}
decreases and assumes a nearly constant value, roughly in the
range 3 - 5 meV. The magnetic field dependence of \emph{U} and
their saturation values are similar to those found in Ru-1212
compound\cite{28} and Bi$_2$Sr$_2$CaCu$_2$O$_8$(BSCCO-2212)
system.\cite{29} However, it is important to notice that values of
\emph{U} in the range  5 - 30 meV are much lower than the ones
found for the less anisotropic compound YBCO which is close to 100
meV.\cite{30}

From the inset of Fig. 6, one observes that the magnetic field
dependence of the activation energy \emph{U} can be described by a
power-law behavior that can be written as \emph{U} $\sim$
\emph{H}$^{-\beta}$. From the fitting parameters we have estimated
$\beta$ $\sim$ 0.33 and 0.32 for \emph{x} = 0.00 and \emph{x} =
0.05, respectively. Values of $\beta$ are of interest because they
reflect the dimensionality of the vortex lattice.\cite{31} For
instance, values of $\beta$ comprehended between 0.33 - 0.5, as
found in the BSCCO-2212 system, indicate a two-dimensional
character of the vortex lattice.\cite{29,30} On the other hand,
$\beta$ $\sim$ 1, as usually found in YBCO cuprates, suggests a
three-dimensional character of the vortice lattice.\cite{32} Thus,
our data strongly indicate that both Ru-1212 and Ir-doped Ru-1212
compounds are very anisotropic and can be classified as having a
vortex lattice with two-dimension character.

The magnetoresistivity data are also useful for an estimate of the
temperature dependence of the upper critical field
\emph{H$_{c2}$(T)} in this series. The temperature dependence of
\emph{H$_{c2}$(T)} of Ru(Ir)-1212 compounds is displayed in Fig.
7, along with the data for the Ru-1222 compound for
comparison.\cite{24} The phase diagram was determined by using the
$\rho$(H,T) curves, considering the same \emph{T$_{c,onset}$} for
all applied magnetic fields and by taking a 50 \% drop of
\emph{$\rho$}(T) as the criterium for the determination of
\emph{H$_{c2}$}. The Ru(Ir)-1212 curves display essentially the
same trend and \emph{H$_{c2}$} shifts to lower temperatures with
increasing Ir content. At high applied magnetic fields, the
upper-critical-field phase diagram shows a linear increase of
\emph{H$_{c2}$} with slopes -3.2 T/K for \emph{x} = 0.00 and -10
T/K for \emph{x} = 0.10. On the other hand, by using the
phenomenological relation

\begin{equation}
\\H_{c2}(T) = H_{c2}(0)\left[1 -
\left(\frac{T}{T_{c}}\right)^{2}\right]^{\alpha},\label{eq3}
\end{equation}

\noindent one can obtain an estimate of \emph{H$_{c2}$}(0) and
\emph{$\alpha$}. The best fitting procedures using Eq. 3 for the
data are plotted in dashed lines in Fig. 7. Fixing the same
\emph{T$_{c}$} used before, the calculated parameters were
\emph{H$_{c2}$}(0) $\sim$ 52 T and $\alpha$ $\sim$ 1.8 for the
sample with \emph{x} = 0.00. The values obtained for all samples
are summarized in Table II. Values of \emph{$\alpha$} in the 1.5 -
2.0 range are frequently observed and are in line with the
reported values for other high-\emph{T$_{c}$} materials.\cite{33}
However, we mention that values of \emph{H$_{c2}$}(0) for pure and
doped compounds are higher than the ones found for Ru-1222. In
addition to this, increasing Ir content has little effect in the
values of \emph{H$_{c2}$}(0).

From previous studies and by considering the values obtained for
the pinning energy it is possible to infer that the highly
anisotropic behavior of Ru-1212 compounds is similar to the
BSCCO-2212 superconductor. The anisotropy factor $\emph{$\gamma$ =
$H_{c2}^{ab}$/H$_{c2}^{c}$}$ of BSCCO-2212 compounds has been
estimated to be in the range between 50 and 200.\cite{31} By
considering a similar anisotropy to Ru-1212 compounds, the
estimated value of \emph{H$_{c2}$} would reflect the upper
critical field parallel to the \emph{a}-\emph{b} plane, and the
superconducting coherence length \emph{$\xi_{c}$(T)} can be
estimated. This can be done by assuming that \emph{H$_{c2}(T)$}
$\simeq$ $H_{c2}^{ab}$(\emph{T}) =
\emph{$\Phi_{0}/2\pi\xi_{c}^{2}$}\emph{(T)}, where \emph{$\Phi$}
is the magnetic flux quantum. The estimated values of
\emph{$\xi_{c}(0)$}$\sim$ 24 $\rm\AA$ are shown in Table II. They
indicate that all the samples studied have similar values of
\emph{$\xi_{c}(0)$}, i.e., that Ir-substitution has little effect
on the coherence length in this series. These results suggest that
the partial substitution of Ru by Ir changes the character of the
Ru-O planes and acts on the coupling between planes. This is
reflected in the nearly constant \emph{H$_{c2}$} for all the
series even when \emph{T$_{c}$} is drastically reduced. Such an
increasing coupling between Cu planes can be also inferred from
the progressive narrowing of the superconducting transition, as
shown in Fig. 4.

\section*{SUMMARY}
The magnetic and transport properties of
Ru$_{1-x}$Ir$_{x}$Sr$_{2}$GdCu$_{2}$O$_{8}$; 0 $\leq$ \emph{x}
$\leq$ 0.10; compounds were investigated. The main results
indicate that Ir substitutes Ru ions with no significant
structural distortions due to similar ionic radii. The substituted
Ir dilutes the Ru magnetic sub-lattice, decreasing both the
magnetic ordering and the superconducting transition temperatures.
The magnetoresistivity data revealed that all samples are
superconducting up to 18 T at temperatures higher than 2 K. On the
other hand the diamagnetic signal in the magnetization curves is
absent for low applied magnetic field. The combined results
indicate that the ruthenocuprates have similar anisotropic
properties as observed in bismuth based high-temperature
superconductors. In addition, the Ir substitutions and the effect
of the applied magnetic field on the electrical resistance curves
suggest that the granular behavior observed may be related to
phase separation of ferromagnetic and antiferromagnetic mesoscopic
regions.

\begin{acknowledgments}
This work was supported by the Brazilian Agency FAPESP under Grant
Nos. 05/53241-9, 01/01455-4, and 01/04231-0, and by the US
National Science Foundation under Grant No. DMR-0306165 (MST).
Work at the NHMFL was performed under the auspices of the NSF, the
State of Florida, and the U.S. Department of Energy. A. R. Jurelo,
F. C. Fonseca, and R. F. Jardim are CNPq fellows under Grant Nos.
150845/2004-9, 301661/2004-9, and 303272/2004-0, respectively.
\end{acknowledgments}

\pagebreak

\pagebreak

{\bf Figure Captions}\\

Figure 1: XRD patterns taken at room temperature of
polycrystalline samples of
Ru$_{1-x}$Ir$_{x}$Sr$_{2}$GdCu$_{2}$O$_{8}$;\emph{x} = 0.00 and
0.10. The figure displays the experimental data (dots), the
calculated diffraction pattern (solid lines), and the difference
between them. The arrow points to the removed 2$\Theta$ region
where the main diffraction peaks of the Sr$_{2}$GdRuO$_{6}$ and
SrRuO$_{3}$ extra phases appear. The inset exhibits the calculated
lattice parameters (\emph{a} and \emph{c}) as a function of the Ir
content.
\\

Figure 2: Temperature dependence of the electrical resistivity in
zero external magnetic field of
Ru$_{1-x}$Ir$_{x}$Sr$_{2}$GdCu$_{2}$O$_{8}$ for Ir concentrations
\emph{x} = 0.00, 0.02, 0.05, and 0.10. The inset displays
\emph{d$\rho$/dT} versus temperature for the compound with
\emph{x} = 0.00 in the neighborhood of \emph{T$_{M}$}. \\

Figure 3: ZFC (open symbols) and FC (full symbols) magnetization
curves as a function of the temperature for
Ru$_{1-x}$Ir$_{x}$Sr$_{2}$GdCu$_{2}$O$_{8}$; 0 $\leq$ \emph{x}
$\leq$ 0.10; measured under the remnant field of the magnet (H
$\sim$ 1 Oe). Upper inset: temperature dependence of the in-phase
component of the $\chi_{ac}$(T) near the magnetic ordering
temperature \emph{T$_{M}$}. Lower inset: plot of the
\emph{T$_{M}$} as a function of the Ir content.
\\

Figure 4: Electrical resistivity versus temperature at several
applied magnetic fields to 18 T for the
Ru$_{1-x}$Ir$_{x}$Sr$_{2}$GdCu$_{2}$O$_{8}$; \emph{x} = 0.00 and
\emph{x} = 0.10; compounds. The remnant field of the
superconducting magnet has been estimated to be 0.03 T.
\\

Figure 5: Temperature dependence of the derivative of the
electrical resistivity for the compound
Ru$_{0.98}$Ir$_{0.02}$Sr$_{2}$GdCu$_{2}$O$_{8}$ in applied
magnetic fields up to 18 T. The upper inset displays a plot of
\emph{d$\rho$/dT} versus temperature at \emph{H} = 14 T in samples
with Ir concentrations \emph{x} = 0.00, \emph{x} = 0.05, and
\emph{x} = 0.10.
\\

Figure 6: Arrhenius plot of the electrical resistivity data for
the compound Ru$_{0.95}$Ir$_{0.05}$Sr$_{2}$GdCu$_{2}$O$_{8}$. The
solid lines represent the best fit to the data by using Eq. (2).
The inset displays a plot of thermal activation energy versus
applied magnetic field for the sample with \emph{x} = 0.00 (open
symbols) and \emph{x} = 0.05 (full symbols).
\\

Figure 7: Temperature dependence of the upper critical field of
Ru-1212, Ir-doped Ru-1212, and Ru-1222. The dashed lines
correspond to the fitting of Eq. 3. The high-field slope
\emph{dH$_{c2}$/dT} is shown in solid lines.
\\

\pagebreak
\begin{table*}
\caption{\label{tabI} Unit cell parameters, atomic parameters, and
agreement factors for Ru$_{1-x}$Ir$_{x}$Sr$_{2}$GdCu$_{2}$O$_{8}$;
\emph{x} = 0.00 and \emph{x} = 0.10; obtained through Rietveld
refinements at room temperature. The space group is
$\emph{P4/mmm}$. The occupancy and the thermal factors were fixed
according to the results of Ref. [8].}
\begin{center}
\begin{tabular}{l l l l l l l}
\hline \hline \ & \emph{a} = \emph{b} (\rm\AA) & \emph{c} (\rm\AA)
& \emph{V}(\rm\AA$^{3}$) & \emph{R$_{wp}$} & \emph{R$_{p}$} &
\emph{$\chi ^{2}$} \\
\hline \ \emph{x} = 0.00 \hspace{0.5cm}  & 3.8389(4)
\hspace{0.5cm} & 11.5652(2) \hspace{0.5cm} & 170.44(3)
\hspace{0.5cm} & 0.0466
\hspace{0.5cm} &  0.0367 & 1.57 \\
\ \emph{x} = 0.10 \hspace{0.5cm}  & 3.8399(1)  \hspace{0.5cm} &
11.5644(5) \hspace{0.5cm} & 170.51(2) \hspace{0.5cm} & 0.0578
\hspace{0.5cm} &  0.0458 & 2.12 \\
\hline
\ Atom & Site & \emph{x} & \emph{y} & \emph{z} & \emph{U} & Occupancy \\
\hline \ Ru   & 1b & 0        & 0   \hspace{0.5cm} & 0.5
\hspace{0.5cm}
& 0.0001 \hspace{0.5cm} & 1(0.9) \\
\ (Ir) &    & 0        & 0   & 0.5       &        & (0.1)  \\
\ Sr   & 2h & 0.5      & 0.5 & 0.3137(3) & 0.0001 & 1      \\
\             &    &          &     & 0.3137(4) &        &        \\
\ Gd   & 1c & 0.5      & 0.5 & 0         & 0.0001 & 1      \\
\ Cu   & 2g & 0        & 0   & 0.1437(4) & 0.0001 & 1      \\
\             &    &          &     & 0.1461(6) &        &        \\
\ O(1)   & 8s & 0.008    & 0   & 0.330     & 0.008  & 0.5    \\
\             &    & 0.007    &     & 0.334     &        &        \\
\ O(2)   & 4i & 0        & 0.5 & 0.125(1)  & 0.008  & 1      \\
\             &    &          &     & 0.122(2)  &        &        \\
\ O(3)   & 4O & 0.138(5) & 0.5 & 0.5       & 0.008   & 0.25   \\
\             &    & 0.158(7) &     &           &        &        \\
\hline \hline
\end{tabular}
\end{center}
\end{table*}

\pagebreak

\begin{table}
\caption{\label{tabII} Several physical parameters extracted from
magnetic and transport properties of
Ru$_{1-x}$Ir$_{x}$Sr$_{2}$GdCu$_{2}$O$_{8}$; 0 $\leq$ \emph{x}
$\leq$ 0.10. The parameters are described in the text and the
corresponding ones belonging to the Ru-1222 compound are also
displayed for comparison.\cite{24} The value of \emph{T$_{c}$} has
been obtained by taking the 50 \% drop of \emph{$\rho(T)$} and
both \emph{H$_{c2}$}(0) and \emph{$\alpha$} from a
phenomenological equation by considering \emph{T$_{c}$} fixed.}
\begin{center}
\begin{tabular}{l l l l l l l}
\hline \hline \ Ir & \emph{T$_{c}$}(K) & \emph{l}(\rm\AA) &
\emph{H$_{c2}$}(0) & \emph{$\alpha$} &
\emph{$\xi_{c}$}(\rm\AA) & \emph{T$_{M}$}(K) \\
\hline
\ 0.00 & 39   & 21 & 52 & 1.8 \hspace{0.5cm} & 25 & 130 \\
\ 0.02 & 38   & 21 & 53 & 1.7 & 24 & 126 \\
\ 0.05 & 32.5 & 15 & 53 & 2.0 & 24 & 120 \\
\ 0.10 & 22   & 15 & 54 & 2.8 & 24 & 112 \\
\hline
\ Ru-1222 & 39 & 58 & 39 & 1.8 & 28 & 100 \\
\hline \hline
\end{tabular}
\end{center}
\end{table}

\pagebreak

\begin{figure}[ht] \centering
\includegraphics[width=10cm]{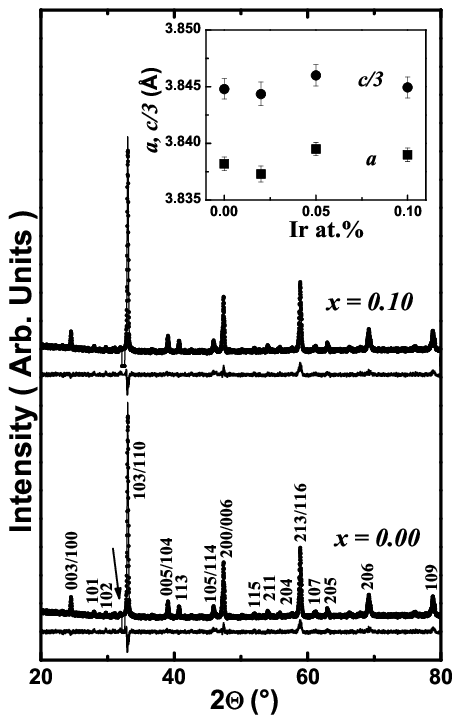}
\caption{\label{Fig1} A. R. Jurelo et al.}
\end{figure}

\pagebreak
\begin{figure}[htp]
\centering
\includegraphics[width=10cm]{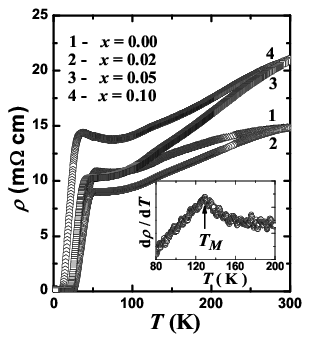}
\caption{\label{Fig2} A. R. Jurelo et al.}
\end{figure}

\pagebreak
\begin{figure}[htp]
\centering
\includegraphics[width=10cm]{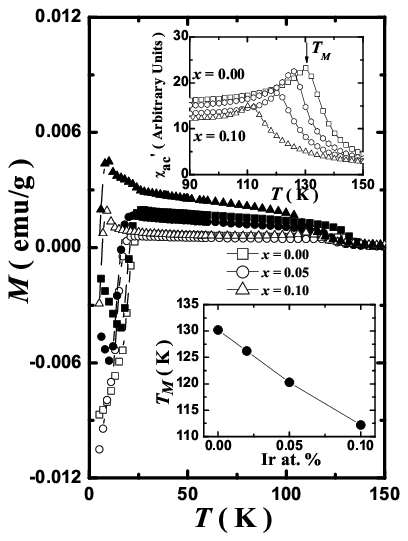}
\caption{\label{Fig3} A. R. Jurelo et al.}
\end{figure}

\pagebreak
\begin{figure}[htp]
\centering
\includegraphics[width=10cm]{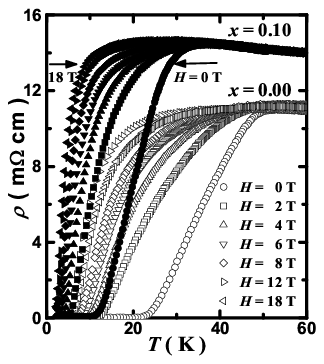}
\caption{\label{Fig4} A. R. Jurelo et al.}
\end{figure}

\pagebreak
\begin{figure}[htp]
\centering
\includegraphics[width=10cm]{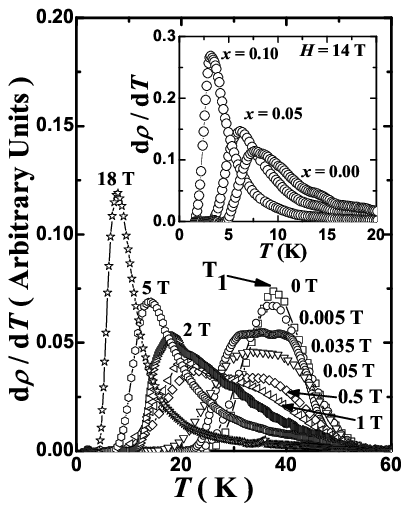}
\caption{\label{Fig5} A. R. Jurelo et al.}
\end{figure}

\pagebreak
\begin{figure}[htp]
\centering
\includegraphics[width=10cm]{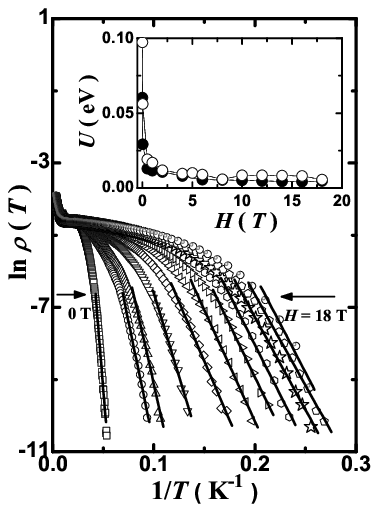}
\caption{\label{Fig6} A. R. Jurelo et al.}
\end{figure}

\pagebreak
\begin{figure}[htp]
\centering
\includegraphics[width=10cm]{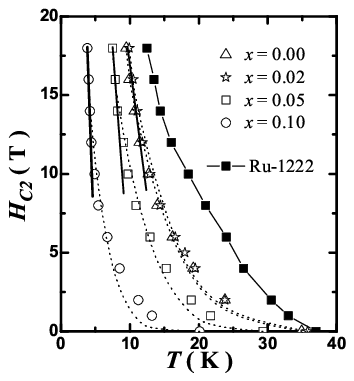}
\caption{\label{Fig7} A. R. Jurelo et al.}
\end{figure}

\end{document}